\documentclass[conference,a4paper]{IEEEtran}

\usepackage{cite}

\usepackage{graphicx,color,epsfig,rotating,subfigure}
\usepackage{amsfonts,amsmath,amssymb}
\usepackage{algorithm,algorithmic}
\usepackage{subfigure}

\long\def\comment#1{}

\newfont{\bbb}{msbm10 scaled 700}

\newfont{\bb}{msbm10 scaled 1100}
\newcommand{\CC}{\mbox{\bb C}}
\newcommand{\PP}{\mbox{\bb P}}

\newcommand{\ZZ}{\mbox{\bb Z}}
\newcommand{\FF}{\mbox{\bb F}}

\newcommand{\EE}{\mbox{\bb E}}

\newcommand{\av}{{\bf a}}

\newcommand{\cv}{{\bf c}}
\newcommand{\dv}{{\bf d}}

\newcommand{\hv}{{\bf h}}

\newcommand{\rv}{{\bf r}}

\newcommand{\tv}{{\bf t}}
\newcommand{\uv}{{\bf u}}
\newcommand{\wv}{{\bf w}}
\newcommand{\vv}{{\bf v}}
\newcommand{\xv}{{\bf x}}
\newcommand{\yv}{{\bf y}}
\newcommand{\zv}{{\bf z}}
\newcommand{\zerov}{{\bf 0}}

\newcommand{\Gm}{{\bf G}}
\newcommand{\Hm}{{\bf H}}

\newcommand{\Tm}{{\bf T}}

\newcommand{\Cc}{{\cal C}}

\newcommand{\Lc}{{\cal L}}

\newcommand{\Nc}{{\cal N}}

\newcommand{\Pc}{{\cal P}}

\newcommand{\Vc}{{\cal V}}

\newcommand{\lambdav}{\hbox{\boldmath$\lambda$}}

\newcommand{\SNR}{{\sf SNR}}
\newcommand{\INR}{{\sf INR}}

\newcommand{\eqdef}{\stackrel{\Delta}{=}}

\newcommand{\transp}{{\sf T}}

\newtheorem{theorem}{Theorem}

\newcommand{\argmin}{\operatornamewithlimits{argmin}}

\begin{document}

\sloppy

%% Paper Title
%% You can use linebreaks \\ within to get better formatting as
%% desired.
\title{Generalized Degrees of Freedom for Network-Coded Cognitive Interference Channel}

%% Author names and affiliations:
%%
%% Avoiding spaces at the end of the author lines is not a problem with
%% conference papers because we don't use \thanks or \IEEEmembership.
%%
%% For several authors with only one affiliation:
%%
% \author{
%   \IEEEauthorblockN{Hui-Ting Chang and Stefan M.~Moser}
%   \IEEEauthorblockA{Department of Electrical and Computer Engineering\\
%     National Chiao Tung University (NCTU)\\
%     Hsinchu, Taiwan\\
%     Email: \{email-of-hui-ting,email-of-stefan\}@ieee.org}
% }
%%
%% For up to three affiliations:
%%
\author{
  \IEEEauthorblockN{Song-Nam Hong}
  \IEEEauthorblockA{Dep. of Electrical Eng.\\
    University of Southern California\\
    Los Angeles, USA\\
    Email: songnamh@usc.edu}
  \and
  \IEEEauthorblockN{Giuseppe Caire}
  \IEEEauthorblockA{Dep. of Electrical Eng.\\
    University of Southern California\\
    Los Angeles, USA\\
    Email: caire@usc.edu}
}
%%
%% For over three affiliations, or if they all won't fit within the width
%% of the page, use this alternative format:
%%
% \author{
%   \IEEEauthorblockN{
%     Michael Shell\IEEEauthorrefmark{1},
%     Homer Simpson\IEEEauthorrefmark{2},
%     James Kirk\IEEEauthorrefmark{3},
%     Montgomery Scott\IEEEauthorrefmark{3} and
%     Eldon Tyrell\IEEEauthorrefmark{4}}
%   \IEEEauthorblockA{
%     \IEEEauthorrefmark{1}School of Electrical and Computer Engineering\\
%     Georgia Institute of Technology, Atlanta, Georgia 30332--0250\\
%     Email: see http://www.michaelshell.org/contact.html}
%   \IEEEauthorblockA{
%     \IEEEauthorrefmark{2}Twentieth Century Fox, Springfield, USA\\
%     Email: homer@thesimpsons.com}
%   \IEEEauthorblockA{
%     \IEEEauthorrefmark{3}Starfleet Academy, San Francisco, California 96678-2391\\
%     Telephone: (800) 555--1212, Fax: (888) 555--1212}
%   \IEEEauthorblockA{
%     \IEEEauthorrefmark{4}Tyrell Inc., 123 Replicant Street, Los Angeles, California 90210--4321}
% }

%% Use for special paper notices

%% To balance the two columns, you should reduce the text-height of
%% the last page using the following command:
%%%%%%%%%%%%%%%%%%%%%%%%%%%%%%%%%%%%%%%%%%%%%%%%%%%%%%%%%%%%%%%%%%%%%
%\addtolength{\textheight}{-9.35cm}
%%%%%%%%%%%%%%%%%%%%%%%%%%%%%%%%%%%%%%%%%%%%%%%%%%%%%%%%%%%%%%%%%%%%%
%% with an appropriate value. This command must be place on the second
%% last page, i.e., for a one-page abstract here, for a two-page
%% abstract right after the \maketitle command.

%% Create the title:
\maketitle

%% Abstract:
%% For the final version of the accepted paper, please make sure you
%% remove the comment "THIS PAPER IS ELIGIBLE FOR THE STUDENT PAPER
%% AWARD."
%%
\begin{abstract}
We study a two-user cognitive interference channel (CIC) where one of the transmitters (primary) has knowledge of a linear combination (over an appropriate finite field) of the two information messages. We refer to this channel model as Network-Coded CIC, since the linear combination may be the result of some linear network coding scheme implemented in the backbone wired network.
%First,  we focus on a finite-field Network-Coded CIC and characterize the capacity region of this channel using
%{\em distributed zero-forcing precoding} for achievability. Then, we extend this scheme using the
%Compute-and-Forward (CoF) framework, and present a novel scheme named Precoded CoF (PCoF) for Gaussian Network-Coded CIC.
In this paper, we characterize the generalized degrees of freedom (GDoF) for the Gaussian Network-Coded CIC.
For achievability, we use the novel Precoded Compute-and-Forward (PCoF) and Dirty Paper Coding (DPC), based on nested lattice codes.
As a consequence of the GDoF characterization, we show that knowing ``mixed data'' (linear combinations of the information messages)
provides a {\em multiplicative} gain for the Gaussian CIC, if the power ratio of signal-to-noise (SNR) to interference-to-noise (INR) is larger than certain threshold. For example, when $\SNR=\INR$, the Network-Coded cognition yields a $100\%$ gain over the classical
Gaussian CIC.
%Finally, numerical results are provided in order to show that the proposed scheme performs well in the range
%of finite SNRs.
\end{abstract}

\section{Introduction}

%Interference is one of the fundamental factors that deteriorate the performance of modern
%communication systems. The two-user interference channel (IC) is a fundamental information-theoretic
%model to study this issue.
%Although the exact characterization of the IC capacity region is a long-standing open problem, even for the Gaussian channel case,
%much progress has been made toward understanding this channel \cite{Kramer,Etkin,Telatar}. Most notably,
%the capacity region of the two-user Gaussian IC was characterized within 1 bit,
%by using superposition coding with an appropriate power allocation of the private and common message codewords,
%and by providing a new upper bounding technique \cite{Etkin}.

\begin{figure}
\centerline{\includegraphics[width=9cm]{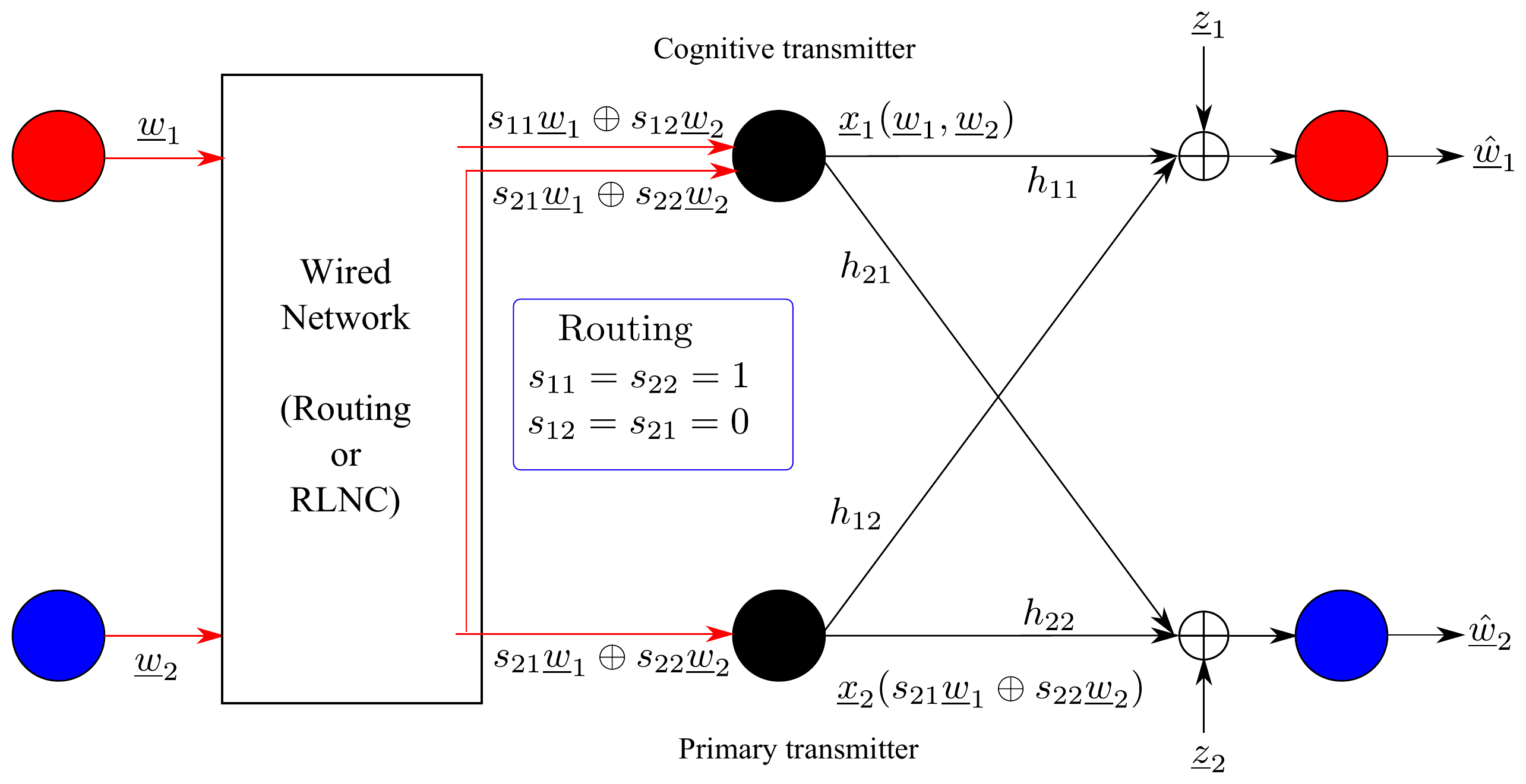}}
\caption{Application of cognitive interference channel (CIC).}
%When routing is used over the wired network, the primary transmitter observes the one of information messages. Yet, when RLNC is used over the wired network, the primary transmitter receives the mixed data (i.e., linear combination of information messages).}
\label{model}
\end{figure}

Transmitters or receivers, in many practical communication systems, are not isolated, and they can share certain
amount of information (i.e., information messages, channel state information, and so on).
For example, in a cloud base station architecture, small base stations (BSs) are spatially distributed
over a certain area, and connected to the infrastructure networks via wired backhaul \cite{Song-IT}. Cooperation among transmitters or receivers can mitigate interferences by forming distributed MIMO systems.
One special case of particular interest is the two-user Cognitive Interference Channel (CIC), where one of the transmitters
(referred to as ``cognitive'') has knowledge of both information messages to the two users, while the other (referred to as ``primary'') has knowledge of the message destined to its intended receiver only.
This model is relevant under certain assumptions on the underlying wired backbone network connecting the two transmitters.
For example, in the case of {\em unidirectional cooperation},  the primary  transmitter sends its message to the
cognitive transmitter via an a wired link of infinite capacity.  Another example is the
asymmetric situation shown in Fig.~\ref{model}, where one transmitter (cognitive) has larger wired backhaul capacity, and therefore
is able to observe both messages.
%Fig.~\ref{model3} is representative of a heterogenous network consisting of a cellular BS
%and home BS (e.g.,  a femtocell access point). The cellular BS is connected to the data router, which generates both messages,
%via a high capacity link supporting rate $2R_{0}$ and the home BS is connected to the same data router via lower capacity link supporting only rate
%$R_{0}$. In this case, the data router sends two information messages to the cellular BS and the one message to the home BS.
The CIC has been extensively investigated in the literature.
The capacity region of the strong interference regime was characterized in \cite{Maric}.
When the interference at the primary receiver is weak, the capacity region was characterized in \cite{Wu,Jovicic,Rini}.
Recently, the capacity region for Gaussian CIC was approximately characterized within 1.87 bits,
regardless of channel parameters \cite{Rini}.

%
%\begin{figure}
%\centerline{\includegraphics[width=10cm]{Figure-Router}}
%\caption{Heterogenous network model as an application of cognitive interference channel (CIC). In the classical CIC, the data router sends the one of information messages to the primary transmitter (i.e., $L(\underline{\wv}_{1},\underline{\wv}_{2}) = \underline{\wv}_{2}$). However, in the Network-Coded CIC, the data router forwards the ``mixed data" to the primary transmitter
%(i.e., $L(\underline{\wv}_{1},\underline{\wv}_{2})=\underline{\wv}_{1} \oplus \underline{\wv}_{2}$).}
%\label{model3}
%\end{figure}

For wired networks, routing is generally optimal only for a single source, multiple intermediate nodes, and a single destination \cite{Ford}.
Yet, it cannot achieve the maximum throughput in the more general case of multiple sources and multiple destinations (multi-source multicasting).
In this case, it is well-known that by allowing intermediate nodes to forward functions of the incoming messages (Network Coding),
the capacity of multi-source multicasting relay networks can be achieved and coincides with the min-cut max-flow bound \cite{Ahlswede}.
Random linear network coding (RLNC) is of particular interest for its practical simplicity. In this case,
intermediate nodes forward liner combinations of the incoming messages by randomly and independently choosing the coefficients
from an appropriate finite-field \cite{Ho}.
Assuming that RLNC is used over the wired network,  in this paper we introduce the Network-Coded CIC as a generalization of the classical CIC,
where the primary transmitter knows a linear combination of the information messages (referred to as ``mixed data''). This
is motivated in Fig.~\ref{model} by introducing RLNC instead of just routing in the backbone network.
%Also, notice that this idea
%is particularly relevant to the heterogenous network model of Fig.~\ref{model3}, where the data router provides mixed data
%(i.e., $\underline{\wv}_{1}\oplus\underline{\wv}_{2}$) to the home BS without violating the backhaul capacity constraint of $R_0$.
Since delivering mixed data at the primary transmitter has the same cost (in terms of backhaul capacity) than delivering a single
message, a natural question arises: {\em Does mixed data at the primary transmitter provide capacity increase ``for free''
for cognitive interference channel?}

\begin{figure}
\centerline{\includegraphics[width=9cm]{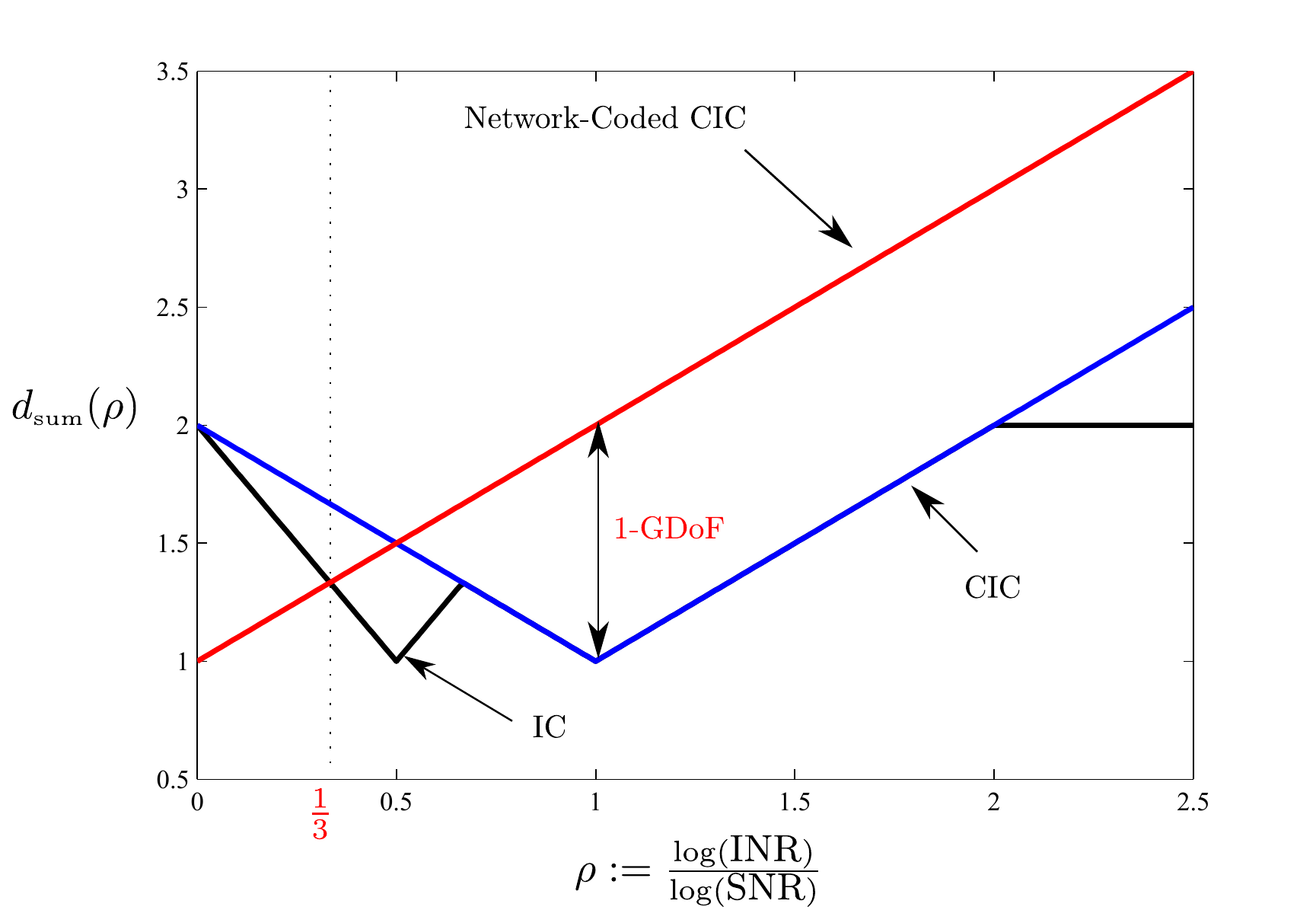}}
\caption{The generalized degrees-of-freedom (GDoF) of the two-user Gaussian Network-Coded cognitive interference channel (CIC).}
%For the interference regimes with $\rho \geq 1/2$, the gap between the Network-Coded CIC and CIC becomes arbitrarily large as SNR and INR goes to infinity. This shows that mixed data at the primary transmitter can provide the unbounded gain.}
\label{GDoF}
\end{figure}

Our main contribution is to approximately characterize the sum capacity of Gaussian Network-Coded CIC in terms of the sum Generalized Degrees of Freedom (GDoF) \cite{Etkin} of the Gaussian Network Coded CIC.
This is enabled by properly using a novel Aligned Precoded Compute-and-Forward (PCoF) and Dirty Paper Coding (DPC).
As a consequence of the GDoF analysis, we show that Network-Coded cognition  can provide {\em multiplicative} gain in cognitive
interference channels.
%The sum GDoF is defined as
%\begin{equation}
%d_{\mbox{\tiny{sum}}}(\rho) = \lim_{\SNR, \INR \rightarrow \infty} \frac{C_{\mbox{\tiny{sum}}}}{\log \SNR},
%\end{equation}
%where $\rho = \log \INR/\log \SNR$, $\SNR$ denotes the signal-to-noise ratio for the direct links and $\INR$ denotes the interference-to-noise ratio
%of the cross links.
As shown in Fig.~\ref{GDoF}, the gain of Network-Coded cognition becomes arbitrary large as
SNR and INR go to infinity as long as $\rho \geq 1/2$ (i.e., except the weak interference regime).
Namely, if $\rho \geq 1/2$,
the performance gap between the Network-Coded CIC and the classical CIC becomes
unbounded.
For example, when $\rho=1$ (i.e., $\SNR = \INR$), Network-Coded cognition provides $100\%$ gain over classical cognition.

%As an application to the heterogenous network of Fig.~\ref{model3},  if the data router has knowledge of the
%dB power ratio of direct-link to cross-link (i.e., $\rho$),  it can send $\underline{\wv}_{2}$ if $\rho \leq 1/2$ and
%$\underline{\wv}_{1} \oplus \underline{\wv}_{2}$, if $\rho > 1/2$.  In this case, the sum GDoF achieved is given by
%\begin{eqnarray}
%d_{\mbox{\tiny{sum}}}(\rho) = \left\{
%                                \begin{array}{ll}
%                                  2-\rho, & \rho < 1/2 \\
%                                  1+\rho, & \rho \geq 1/2.
%                                \end{array}
%                              \right.
%\end{eqnarray}
%This paper is organized as follows.
%In Section \ref{sec:pre} we define the system model, summarize some definitions on lattices and lattice coding,
%and review CoF. In Section \ref{sec:PCoF} we characterize the capacity region of finite-field Network-Coded CIC
%and present the Aligned PCoF, as a natural extension of finite-field scheme, for Gaussian Network-Coded CIC.
%In section \ref{sec:DPC}, we derive an achievable rate region of Gaussian Network-Coded CIC, based on Aligned PCoF
%and DPC. We characterize the sum GDoF the Gaussian Network-Coded CIC in Section \ref{sec:GDoF}.
%Some concluding remarks are provided in Section \ref{sec:conclusion}.

%%%%%%%%%%%%%%%%%%%%%%%%%%%%%%%%%%%%%%%%%%%%%%%%%%%%%%%%%%%%%%%%%%%%%%%%%%%%%%%%%%%%%%%%%%%%%%%%%%%
\section{Preliminaries}\label{sec:pre}

In this section we provide some basic definitions and results which will be extensively used in the sequel.
%%%%%%%%%%%%%%%%%%%%%%%%%%%%%%%%%%%%%%%%%%%%%%%%%%%%%%%%%%%%%%%%%%
\subsection{System Model}

A two-user Gaussian Network-Coded CIC consists of a Gaussian interference channel where
transmitter 1 (the cognitive transmitter) knows both user 1 and user 2 information messages (or, equivalently, two linearly independent linear combinations thereof) and transmitter 2 (the primary transmitter) only knows only one linear combination of the messages.
Without loss of generality, we assume that the cognitive transmitter knows
($\underline{\wv}_{1},\underline{\wv}_{2})$, and the primary transmitter has
$\underline{\wv}_{1} \oplus \underline{\wv}_{2}$, where $\underline{\wv}_{\ell} \in \FF_{q}^{r}$ denotes
the information message desired at receiver $\ell$, at rate $R_\ell$ bit/symbol, for $\ell=1,2$.
We assume that if $R_{1}  \neq R_{2}$ then the lowest rate message is zero-padded
such that both messages have a common length,  given by $r = \max\{nR_{1},nR_{2}\}$, where $n$ denotes the
coding block length.  A block of $n$ channel uses of the discrete-time complex baseband two-user IC is described by
\begin{eqnarray}
\underline{\yv}_{1} &=& h_{11} \underline{\xv}_{1} + h_{12} \underline{\xv}_{2} + \underline{\zv}_{1}\label{eq:channel1}\\
\underline{\yv}_{2} &=& h_{21} \underline{\xv}_{1} + h_{22} \underline{\xv}_{2} + \underline{\zv}_{2}, \label{eq:channel2}
\end{eqnarray}
where $\underline{\zv}_{\ell} \in \CC^{n \times 1}$ contains i.i.d. Gaussian noise samples $\sim \Cc\Nc(0,1)$
and $h_{ij} \in \CC$ denotes the channel coefficients, assumed to be constant over the whole block of length $n$ and known to all nodes.
Also, we have a power constraint, given by $\frac{1}{n} \EE[\|\underline{\xv}_{\ell}\|^2]\leq \SNR$ for $\ell=1,2$, where $\|\cdot\|$ denotes
the $\ell_{2}$-norm. Each receiver $\ell$ observes the channel output $\underline{\yv}_{\ell}$ and produces an estimate $\hat{\underline{\wv}}_{\ell}$
of the desired message $\underline{\wv}_{\ell}$. We say that receiver $\ell$ is in error whenever $\hat{\underline{\wv}}_{\ell} \neq \underline{\wv}_{\ell}$. A rate pair $(R_{1},R_{2})$ is achievable if there exists a family of codebooks with codewords satisfying the power constraint,
and corresponding decoding functions such that the average decoding error probability satisfies
$\lim_{n \rightarrow \infty}\PP(\hat{\underline{\wv}}_{\ell} \neq \underline{\wv}_{\ell}) = 0$, for $\ell=1,2$.

%%%%%%%%%%%%%%%%%%%%%%%%%%%%%%%%%%%%%%%%%%%%%%%%%%%%%%
\subsection{Nested Lattice Codes}\label{subsec:NLC}

Let $\ZZ[j]$ be the ring of Gaussian integers and $p$ be a prime. Let $\oplus$ denote the addition over $\FF_{q}$ with $q=p^2$, and let $g: \FF_{q} \rightarrow \CC$ be the natural mapping of $\FF_{q}$ onto $\{a+jb: a,b \in \ZZ_{p}\} \subset \CC$.
We recall the nested lattice code construction given in \cite{Nazer2011}.
Let $\Lambda = \{ \underline{\lambdav} = \underline{\zv} \Tm : \underline{\zv} \in \ZZ^n[j]\}$ be a lattice in $\CC^n$,
with full-rank generator matrix $\Tm \in \CC^{n \times n}$.
Let $\Cc  = \{ \underline{\cv} = \underline{\wv} \Gm : \wv \in \FF_{q}^r \}$ denote a linear code over $\FF_{q}$ with block length $n$ and dimension $r$, with generator matrix $\Gm$. The lattice $\Lambda_1$ is defined through ``construction A'' (see \cite{Erez2004} and references therein) as
\begin{equation} \label{construction-A}
\Lambda_1 = p^{-1} g(\Cc) \Tm + \Lambda,
\end{equation}
where $g(\Cc)$ is the image of $\Cc$ under the mapping $g$ (applied component-wise).
It follows that $\Lambda \subseteq \Lambda_1 \subseteq p^{-1} \Lambda$ is a chain of nested lattices, such that
$|\Lambda_1/\Lambda| = p^{2r}$ and $|p^{-1} \Lambda/\Lambda_1| = p^{2(n - r)}$.

For a lattice $\Lambda$ and $\underline{\rv} \in \CC^n$,
we define the lattice quantizer $Q_{\Lambda}(\underline{\rv}) = \argmin_{\underline{\lambdav} \in \Lambda}\|\underline{\rv} - \underline{\lambdav} \|^2$,
the Voronoi region $\Vc_\Lambda = \{\underline{\rv} \in \CC^{n}: Q_{\Lambda}(\underline{\rv}) = \underline{\zerov}\}$
and $[\underline{\rv}] \mod \Lambda = \underline{\rv} - Q_{\Lambda}(\underline{\rv})$.
For $\Lambda$ and $\Lambda_1$ given above, we define the lattice code
$\Lc = \Lambda_{1} \cap \Vc_\Lambda$ with rate $R = \frac{1}{n} \log |\Lc| = \frac{r}{n}\log{q}$.
Construction A provides a {\em natural labeling}
of the codewords of $\Lc$ by the information messages $\underline{\wv} \in \FF^r_{q}$.  Notice that the set $p^{-1} g(\Cc)\Tm$ is a {\em system of coset representatives}
of the cosets  of $\Lambda$ in $\Lambda_1$. Hence, the natural labeling function  $f : \FF^r_{q} \rightarrow \Lc$
is defined by $f(\underline{\wv}) = p^{-1} g(\underline{\wv} \Gm)\Tm \mod \Lambda$.

%%%%%%%%%%%%%%%%%%%%%%%%%%%%%%%%%%%%%%%%%%%%%%%%%%
\subsection{Compute-and-Forward} \label{subsec:CoF}

We recall here the CoF scheme of \cite{Nazer2011}. Consider the two-user Gaussian multiple access channel defined by
\begin{equation} \label{GMAC}
\underline{\yv} = \sum_{k = 1}^{2} h_{k} \underline{\xv}_{k} + \underline{\zv},
\end{equation}
where $\hv = [h_{1},h_{2}]^\transp$ and the elements of $\underline{\zv}$ are i.i.d. $\sim \Cc\Nc(0,1)$.
All users make use of the same nested lattice codebook $\Lc = \Lambda_1 \cap \Vc_\Lambda$,
where $\Lambda$ has {\em second moment} $\sigma_\Lambda^2 \eqdef \frac{1}{n \mbox{Vol}(\Vc)} \int_{\Vc} \| \underline{\rv} \|^2 d\underline{\rv} = \SNR$. Each user $k$ encodes its information message $\underline{\wv}_{k} \in \FF_{q}^{r}$ into the corresponding codeword $\underline{\tv}_{k} = f(\underline{\wv}_k)$
and produces its channel input according to
\begin{equation}\label{eq:channelinput}
\underline{\xv}_{k} = \left [\underline{\tv}_{k} + \underline{\dv}_{k} \right ] \mod \Lambda,
\end{equation}
where the {\em dithering sequences} $\underline{\dv}_{k}$'s are mutually independent across the users, uniformly distributed
over $\Vc_\Lambda$, and known to the receiver.  The decoder's goal is to recover a
linear combination $\underline{\vv} = [\sum_{k =1}^{2} a_{k} \underline{\tv}_{k}] \mod \Lambda$ with {\em integer coefficient vector}
$\av = [a_1, a_2]^\transp \in \ZZ^2[j]$.
Since $\Lambda_1$ is a $\ZZ[j]$-module (closed under linear combinations with Gaussian integer coefficients),
then $\underline{\vv} \in \Lc$.  Letting $\hat{\underline{\vv}}$ be the decoded codeword (for some decoding function which
in general depends on $\hv$ and $\av$), we say that a computation rate $R$ is achievable for this setting if there exists sequences
of lattice codes  $\Lc$ of rate $R$ and increasing block length $n$, such that the decoding error probability satisfies
$\lim_{n \rightarrow \infty} \PP(\hat{\underline{\vv}} \neq \underline{\vv} ) = 0$.

In the scheme of \cite{Nazer2011}, the receiver computes
\begin{eqnarray}
\hat{\underline{\yv}} & = & \left [\alpha \underline{\yv} - \sum_{k=1}^{2}a_{k} \underline{\dv}_{k} \right ] \mod \Lambda \nonumber \\
&=& \left [\underline{\vv} + \underline{\zv}_{\mbox{\tiny{eff}}} (\hv, \av,\alpha) \right ] \mod \Lambda, \label{eq:received}
\end{eqnarray} where
\begin{equation}
\underline{\zv}_{\mbox{\tiny{eff}}} (\hv, \av, \alpha) = \sum_{k=1}^{2} (\alpha h_{k} - a_{k}) \underline{\xv}_{k} + \alpha \underline{\zv} \label{eq:effnoise}
\end{equation}
denotes the {\em effective noise}, including the non-integer self-interference (due to the fact that $\alpha h_k \notin \ZZ[j]$ in general) and the additive Gaussian noise term.  The scaling, dither removal and modulo-$\Lambda$ operation in (\ref{eq:received})
is referred to as the {\em CoF receiver mapping} in the following.
%By minimizing the variance of $\underline{\zv}_{\mbox{\tiny{eff}}}(\hv,\av,\alpha)$
%with respect to $\alpha$, we obtain
%\begin{eqnarray}
%\sigma^{2} (\hv,\av) &=& \min_{\alpha} \sigma_{z_{\mbox{\tiny{eff}}}}^{2}(\hv,\av,\alpha) \nonumber \\
%&=& \av^{\herm}(\SNR^{-1}\Id+\hv\hv^{\herm})^{-1}\av\label{eq:effnoise1}.
%\end{eqnarray}
%Since $\alpha$ is uniquely determined by $\hv$ and $\av$, it will be omitted in the following,
%for the sake of notation simplicity.
From \cite{Nazer2011}, we know that by applying lattice decoding to $\hat{\underline{\yv}}$ given in (\ref{eq:received}) the following computation rate is achievable:
\begin{equation}\label{eq:CoFrate}
R(\hv,\alpha,\av,\SNR) = \log^{+}\Big(\frac{\SNR}{\|\alpha\|^2+\|\alpha\hv-\av\|^2\SNR}\Big),
\end{equation}
where $\log^{+}(x) \triangleq \max\{\log(x),0\}$.
\section{An achievable rate region for Gaussian Network-Coded CIC}\label{sec:DPC}

%It was shown in Section \ref{sec:PCoF} that the distributed zero-forcing precoding is optimal for finite-field Network-Coded CIC. In Gaussian channel, however, the channel coefficients are not integers and hence Aligned PCoF may not be optimal due to the non-integer penalty.
Using the fact that the cognitive transmitter has non-causal information of the primary transmitter signal, it can totally eliminate  the known interference at its own intended receiver by using DPC. Also, we can remove the interference at the receiver 2 using Aligned PCoF. Using CoF decoding, the receiver 2 can reliably decode an integer linear combination of the lattice codewords sent by transmitters. The ``interference" in the finite-field domain can be completely eliminated by precoding over the finite-field at the cognitive transmitter. It is well known that the performance of CoF is deteriorated by the non-integer penalty (i.e., the residual ``self-interference" due to the fact that the channel coefficients take on non-integer values in practice). In order to eliminate this penalty, the primary transmitter scales its signal by some constant $\beta \in \Pc$ to create more favorable channel for CoF receiver mapping where $\Pc=\{\beta \in \CC:\|\beta\|\leq 1\}$.

%\begin{figure*}
%\centerline{\includegraphics[width=16cm]{FIG-DECODING}}
%\caption{Encoding and decoding structures of proposed scheme. The cognitive transmitter performs the DPC and the primary transmitter performs the PCoF.}
%\label{decoding}
%\end{figure*}

We let $\av=[a_{1},a_{2}] \in \ZZ[j]^{2}$ denote the integer coefficients vector used at receiver 2 for the modulo-$\Lambda$ receiver mapping (\ref{eq:received}), and we let $q_{\ell} = g^{-1}(a_{\ell} \mod p\ZZ[j])$.
For the time being, it is assumed that  $q_{1},q_{2} \neq 0$ over $\FF_q$.
The proposed achievability scheme proceeds as follows
%(see Fig.~\ref{decoding}):
\begin{itemize}
\item The primary transmitter produces the lattice codeword $\underline{\vv}_{2}= f(\underline{\wv}_{1} \oplus \underline{\wv}_{2})$ and transmits the following channel inputs:
\begin{equation}
 \underline{\xv}_{2} = [\underline{\vv}_{2} + \underline{\dv}_{2}] \mod \Lambda.
\end{equation}
\item The cognitive transmitter produces the precoded message $b\underline{\wv}_{1}$ where $b \in \FF_{q}$ is given by
\begin{equation}\label{eq:con}
q_{1}b \oplus q_{2} = 0 \Rightarrow b = (q_{1})^{-1}(-q_{2})
\end{equation}
where $(q_{1})^{-1}$ denotes the multiplicative inverse of $q_{1}$ and $(-q_{2})$ denotes the additive inverse of $q_{2}$.
\item The cognitive transmitter performs DPC using the known interference signal $h_{12}\underline{\xv}_{2}$ to get:
\begin{equation}
\underline{\xv}_{1} = [\underline{\vv}_{1} - \alpha_{1}(h_{12}/h_{11})\underline{\xv}_{2} + \underline{\dv}_{1}] \mod \Lambda,
\end{equation}
where $\underline{\vv}_{1} = f(b\underline{\wv}_{1})$. The known interference signal is in fact generated
by using the knowledge of the message $\underline{\wv}_{1}\oplus \underline{\wv}_{2}$,
the dense lattice codebooks, and the dithering sequence $\underline{\dv}_{2}$ %(see Fig.~\ref{decoding}).
The $\dv_{\ell}$'s are mutually independent across the transmitters, uniformly distributed over $\Vc_{\Lambda}$, and known to all nodes.
\end{itemize}
Because of linearity, the precoding and the encoding over the finite-field commute. Therefore, we can
write
\begin{eqnarray}
\underline{\vv}_{1} &=& g(b)\underline{\tv}_{1} \mod \Lambda\\
\underline{\vv}_{2} &=& \underline{\tv}_{1} + \underline{\tv}_{2} \mod \Lambda
\end{eqnarray} where $\underline{\tv}_{1} = f(\underline{\wv}_{1})$ and $\underline{\tv}_{2} = f(\underline{\wv}_{2})$.
%Receivers 1 and 2 observe the $\underline{\yv}_{1}$ and $\underline{\yv}_{2}$:
%\begin{eqnarray}
%\underline{\yv}_{1} &=& h_{11}\underline{\xv}_{1} + h_{12}\underline{\xv}_{2} + \underline{\zv}_{1}\\
%\underline{\yv}_{2} &=& h_{21}\underline{\xv}_{1} + h_{22}\underline{\xv}_{2} + \underline{\zv}_{2}.
%\end{eqnarray}
Receiver 1 performs the inflated modulo-lattice mapping as $\hat{\underline{\yv}}_{1} = [\alpha_{1}\underline{\yv}_{1}/h_{11} - \underline{\dv}_{1}] \mod \Lambda$.
Then the resulting channel is a mod-$\Lambda$ additive noise channel as
\begin{eqnarray*}
\hat{\underline{\yv}}_{1}
%&=& [\alpha_{1} [h_{11}\underline{\xv}_{1} + h_{12} \underline{\xv}_{2} + \underline{\zv}_{1}] -\underline{\dv}_{1}]\mod \Lambda \\
%&=& [\alpha_{1} [h_{11}\underline{\vv}_{1} - \alpha_{1}h_{11}(h_{12}\underline{\xv}_{2})\\
%&& + h_{11}\underline{\dv}_{1} + h_{12}\underline{\xv}_{2} + \underline{\zv}_{1}] - \underline{\dv}_{1}] \mod \Lambda \\
%&=& [\underline{\vv}_{1} -(1-\alpha_{1}h_{11})\underline{\vv}_{1} +\alpha_{1}(1-\alpha_{1}h_{11})h_{12}\underline{\xv}_{2} \\
%&&-(1-\alpha_{1}h_{11})\underline{\dv}_{1} +\alpha_{1}\underline{\zv}_{1}] \mod \Lambda\\
&=& [\underline{\vv}_{1} -(1-\alpha_{1}h_{11})\underline{\uv}_{1} + \alpha_{1}\underline{\zv}_{1}/h_{11} ] \mod \Lambda
\end{eqnarray*} where  $\underline{\uv}_{1}$ is uniformly distributed on $\Vc_{\Lambda}$ and is independent of $\underline{\zv}_{1}$ and $\underline{\vv}_{1}$ by Crypto Lemma. From standard DPC results \cite{Zamir}, choosing
\begin{equation}
\alpha_{1}=\alpha_{1,\mbox{\tiny{MMSE}}} \eqdef \frac{\SNR \|h_{11}\|^2}{1 + \SNR \|h_{11}\|^2},
\end{equation} we obtain
\begin{equation}
R_{1} \leq \log(1+\|h_{11}\|^2\SNR).
\end{equation}
Letting $\tilde{\hv}=[h_{21},\tilde{h}_{22}]$ with $\tilde{h}_{22}=h_{22} - \alpha_{1,\mbox{\tiny{MMSE}}}h_{12}h_{21}/h_{11}$, receiver 2 applies the CoF receiver mapping in (\ref{eq:received}) with integer coefficients vector
 $\av=(a_{1},a_{2}) \in \ZZ[j]^{2}$ and scaling factor $\alpha_{2} = a_{1}/h_{21}$, yielding
\begin{eqnarray*}
\hat{\underline{\yv}}_{2} &=& [a_{1}\underline{\vv}_{1} + a_{2} \underline{\vv}_{2} +\alpha_{2}(h_{21}\underline{\xv}_{1}+h_{22}\underline{\xv}_{2} + \underline{\zv}_{2})\\
&& - a_{1}[\underline{\vv}_{1}+\underline{\dv}_{1}]-a_{2}[\underline{\vv}_{2}+\underline{\dv}_{2}]] \mod \Lambda\\
%&=&[a_{1}\underline{\vv}_{1}+a_{2}\underline{\vv}_{2}+\alpha_{2}h_{21}[\underline{\vv}_{1}-\alpha_{1,\mbox{\tiny{MMSE}}}(h_{12}/h_{11})\underline{\xv}_{2}\\
%&&+\underline{\dv}_{1}+\underline{\lambdav}] + \alpha_{2}h_{22}\underline{\xv}_{2}+\alpha_{2}\underline{\zv}_{2} - a_{1}[\underline{\vv}_{1}+\underline{\dv}_{1}]\\
%&&-a_{2}\underline{\xv}_{2}] \mod  \Lambda \\
&=&[a_{1}\underline{\vv}_{1} + a_{2} \underline{\vv}_{2} + (\alpha_{2}h_{21}-a_{1})[\underline{\vv}_{1}+\underline{\dv}_{1}] \\
&&+ (\alpha_{2}\tilde{h}_{22} - a_{2})\underline{\xv}_{2} +\alpha_{2}h_{21}\underline{\lambdav} + \alpha_{2}\underline{\zv}_{2}] \mod \Lambda\\
&\stackrel{(a)}{=}& \left[\av^{\transp}\left[
                                                 \begin{array}{c}
                                                   \underline{\vv}_{1} \\
                                                   \underline{\vv}_{2} \\
                                                 \end{array}
                                               \right]+(\alpha_{2}\tilde{h}_{22}-a_{2})\underline{\uv}_{2}+\alpha_{2}\underline{\zv}_{2}\right] \mod \Lambda\\
&=&\Big[\left(\av^{\transp}\left[
                        \begin{array}{cc}
                          g(b) & 0 \\
                          1 & 1 \\
                        \end{array}
                      \right] \mod p\ZZ[j]\right)
\left[
                                                 \begin{array}{c}
                                                   \underline{\tv}_{1} \\
                                                   \underline{\tv}_{2} \\
                                                 \end{array}
                                               \right]\\
&& + \underline{\zv}_{\mbox{\tiny{eff}}}(\tilde{\hv},\av)
\Big]  \mod \Lambda\\
&\stackrel{(b)}{=}& [([a_{2}] \mod p\ZZ[j])\underline{\tv}_{2} +\underline{\zv}_{\mbox{\tiny{eff}}}(\tilde{\hv},\av)] \mod \Lambda                               
\end{eqnarray*} where $\underline{\lambdav}=Q_{\Lambda}(\underline{\vv}_{1}-\alpha_{1,\mbox{\tiny{MMSE}}}(h_{12}/h_{11})\underline{\xv}_{2}+\underline{\dv}_{1})$, 
$(a)$ is due to the fact that $\alpha_{2}h_{21}\underline{\lambdav}= a_{1}\underline{\lambdav} \in \Lambda$, and  $(b)$ follows from the fact that the $b$ is chosen to satisfy the (\ref{eq:con}), i.e., $a_{1}g(b) + a_{2} \mod p\ZZ[j] = 0$, and
\begin{equation}
 \underline{\zv}_{\mbox{\tiny{eff}}}(\tilde{\hv},\av) = (a_{1}\tilde{h}_{22}/h_{21} - a_{2})\underline{\uv}_{2} + (a_{1}/h_{21})\underline{\zv}_{2}.\label{eq:enoise}
\end{equation} By applying the lattice coding to $\hat{\underline{\yv}}_{2}$, the receiver 2 can decode its message if
\begin{equation}
R_{2} \leq R(\tilde{\hv},a_{1}/h_{21},\av,\SNR).% \log\left(\frac{\SNR}{\|a_{1}/h_{21}\|^2 +\|a_{1}\tilde{h}_{22}/h_{21}-a_{2}\|^2\SNR}\right).
\end{equation} In order to mitigate the non-integer penalty at receiver 2,  the primary transmitter only scales its signal by some constant $\beta \in \Pc$. In this way,  the rate $R_{1}$ in (\ref{eq:rate1}) is preserved, and the rate $R_{2}$ can be rewritten as a function of $\beta \in \Pc$ as:
\begin{equation}
R_{2}(\beta) \leq R(\tilde{\hv}(\beta),a_{1}/h_{21},\av,\SNR), \label{eq:PCoF}
\end{equation}
where now we have
\begin{equation} \label{htilde}
\tilde{\hv}(\beta) = \left [ h_{21}, \beta \left ( h_{22} - \frac{\SNR\|h_{11}\|^2}{1+\|h_{11}\|^2\SNR} h_{12} h_{21}/h_{11} \right ) \right ],
\end{equation}
for some $\beta \in \Pc$. Hence, we have proved the following:

\begin{theorem}\label{thm:DPC} Aligned PCoF and DPC applied to Gaussian Network-Coded CIC with $\Hm=[h_{ij}] \in \CC^{2 \times 2}$  achieves the rate pairs $(R_{1},R_{2})$ such that
\begin{eqnarray}
R_{1} &\leq& \log(1+\|h_{11}\|^2\SNR)\label{eq:rate1}\\
R_{2} &\leq&  R(\tilde{\hv},a_{1}/h_{21},\av,\SNR)
\end{eqnarray} for any $\av \in \ZZ[j]^{2}$ with $a_{1},a_{2} \neq 0$ and any $\beta \in \Pc$, where $\tilde{\hv}$ is given in (\ref{htilde}).
\hfill \IEEEQED
\end{theorem}
%
%\begin{example} We evaluate the performance of proposed schemes with respect to their average achievable sum rates. We computed the {\em ergodic} sum rates by Monte Carlo averaging with respect to the channel realizations with $h_{ij} \sim \Cc\Nc(0,1)$. Also, we considered the performance of full-cooperation (i.e., vector broadcast channel with sum-power constraint) as a simple upper bound. The capacity of this channel can be computed using the efficient algorithm provided in \cite{Hoon}, based on Lagrangian duality.
%In Fig.~\ref{simulation}, Aligned PCoF shows the satisfactory performance in the moderate SNRs (i.e., $\SNR < 20$ dB). Yet, this scheme seems to suffer from the non-integer penalty at high SNRs, having a larger gap with Aligned PCoF and DPC. It is noticeable that Aligned PCoF and DPC almost achieves the performance of full-cooperation within a constant gap.
%\end{example}

%\begin{figure}
%\centerline{\includegraphics[width=12cm]{simulation_fading}}
%\caption{Average sum rates for Gaussian Network-Coded CIC with channel coefficients $\sim \Cc\Nc(0,1)$.}
%\label{simulation}
%\end{figure}

%%%%%%%%%%%%%%%%%%%%%%%%%%%%%%%%%%
\section{Generalized Degrees of Freedom}\label{sec:GDoF}

In the high SNR regime, a useful proxy for the performance of wireless networks
is provided by the sum Degree-of-Freedom (DoF), which is the pre-log factor (multiplexing gain)
in the expression of the  sum capacity in terms of SNR.  In this section we study the symmetric Generalized DoF (GDoF)
as introduced in \cite{Etkin}, which is a more refined proxy for the high-SNR performance, capturing the relative strength
of direct and interference links. We consider the following channel model:
\begin{eqnarray}
\underline{\yv}_{1} &=& h_{11}\sqrt{\SNR} \underline{\xv}_{1} + h_{12}\sqrt{\INR} \underline{\xv}_{2} + \underline{\zv}_{1}\label{eq:mGDoF1}\\
\underline{\yv}_{2} &=& h_{21}\sqrt{\INR} \underline{\xv}_{1} + h_{22}\sqrt{\SNR} \underline{\xv}_{2} + \underline{\zv}_{2}\label{eq:mGDoF2}
\end{eqnarray}where $h_{ij} \in \CC$ are bounded non-zero constants independent of $\SNR, \INR$,
$\underline{\zv}_{\ell}$ is the i.i.d. Gaussian noise $\sim \Cc\Nc(0,1)$, and $\frac{1}{n}\EE[\|\underline{\xv}_{\ell}\|^2] \leq 1$ for $\ell=1,2$.
The channel is parameterized by $\SNR$ and $\INR$, both growing to infinity. The way these parameters grow to infinity if
defined by $\rho > 0$, given by
\begin{equation}
\rho =\log\INR/\log\SNR,
\end{equation}
$i.e., $ by letting $\INR = \SNR^{\rho}$ as $\SNR \rightarrow \infty$. The sum GDoF is defined by
\begin{equation}
d_{\mbox{\tiny{sum}}}(\rho) = \lim_{\SNR \rightarrow \infty} \frac{C_{sum}}{\log\SNR}.
\end{equation}
The main result of this section is given by:
\begin{theorem}\label{thm:GDoF}
For the Gaussian Network-Coded CIC, the sum symmetric GDoF is given by
\begin{eqnarray}
d_{\mbox{\tiny{sum}}}(\rho) = 1+\rho.
\end{eqnarray}
\end{theorem}
\begin{IEEEproof}
See Appendix~\ref{proof:GDoF}.
\end{IEEEproof}

In order to demonstrate the benefit gain of the mixed message at the primary transmitter,
we compare the sum GDoF of Gaussian IC and Gaussian CIC.
The GDoF of Gaussian IC is computed in \cite{Etkin}. Also, from the constant gap result in \cite{Rini}, we can immediately compute the GDoF of Gaussian CIC. The GDoFs of three channels are plotted in Fig.~\ref{GDoF}.

\section{Concluding Remarks}\label{sec:conclusion}

We investigated a two-user cognitive interference channel (CIC),
in the case where the ``primary'' transmitter   knows  a linear combination of the information messages.
%First, we characterized the capacity region of the finite-field Network Coded CIC, based on the
%{\em distributed zero-forcing precoding} technique.
%We extended this scheme to the  Gaussian case  by using the framework of Compute-and-Forward (CoF).
%Our new proposed scheme is referred to as Precoded CoF (PCoF).
%Further, we introduced ``Aligned" PCoF where each transmitter scales its signal by some constant, in order to
%create more favorable channel for the integer-forcing CoF receiver mapping.
%We provided an achievability technique that combines DPC with Aligned PCoF, and argued that this scheme does not suffer from
%integer penalty and is able to achieve the cooperative vector broadcast channel outer bound (full cooperation of transmitters).
The proposed combination of Aligned PCoF and Dirty Paper Coding, based on nested lattice codes, allowed us to
characterize  the sum generalized degrees-of-freedom of the Gaussian Network-Coded CIC. In particular,
our result shows the surprising fact that, in certain regimes of the SNR/INR scaling region,
network-coded cognition yields an unbounded gain (i.e., multiplicative gain) in the Gaussian CIC, with respect to
the classical cognitive transmitter model.

\appendices
\section{Proof of Theorem \ref{thm:GDoF}}\label{proof:GDoF}

\subsection{Achievable scheme}

We use the achievable rates given in Theorem \ref{thm:DPC}. It is immediately shown that the achievable GDoF of the cognitive transmitter is 1, obtained by
\begin{equation}\label{eq:GDoF1}
d_{1}(\rho) = \lim_{\SNR \rightarrow \infty} \frac{\log(1+\|h_{11}\|^2\SNR)}{\log\SNR} = 1.
\end{equation} In this proof, we show that the primary transmitter achieves the $\rho$ GDoF by carefully choosing the $\beta$. The effective channel for Aligned PCoF is given in (\ref{htilde}) as
$\tilde{\hv}(\beta) = [h_{21}\sqrt{\INR}, \beta(h_{22}\sqrt{\SNR}-\alpha_{1,\mbox{\tiny{MMSE}}}(h_{12}h_{21}/h_{11})\SNR^{\rho-\frac{1}{2}})]$ and can be rewritten as
\begin{equation}
\tilde{\hv}(\beta) = \SNR^{\rho/2}[h_{21},\beta\tilde{h}_{22}]
\end{equation} where $\tilde{h}_{22} = h_{22}\SNR^{(1-\rho)/2} - h \SNR^{(\rho - 1)/2}$ and $h=\alpha_{1,\mbox{\tiny{MMSE}}}(h_{12}h_{21}/h_{11})$.
Here, we choose
$\beta=\beta^{\star} \triangleq h_{21}/(\tilde{h}_{22}\gamma)$, where $\gamma \geq 1$ is an integer with
$\gamma=\lceil\|h_{21}/\tilde{h}_{22}\|\rceil \in \ZZ_{+}$.
This produces a kind of ``aligned" channel:
\begin{equation}
\tilde{\hv}= \SNR^{\rho/2}[h_{21},h_{21}/\gamma].
\end{equation}
Letting $a_{1}=\gamma$, and $a_{2}=1$, the effective noise in (\ref{eq:enoise}) is obtained by
\begin{eqnarray}
\zv_{\mbox{\tiny{eff}}}(\tilde{\hv},\av) = (\gamma/(h_{21}\SNR^{\rho/2}))\underline{\zv}_{2}.
\end{eqnarray}This shows that non-integer penalty is completely eliminated. Also, we can use the zero forcing precoding over $\FF_{q}$ since the integer coefficients $a_{1}$ and $a_{2}$ are non-zero. From this, we have the lower bound on the achievable rate of Aligned PCoF:
\begin{equation}
\max_{\beta}R_{2}(\beta) \geq R_{2}(\beta^{*}) = \rho\log(\|h_{21}\|^2\SNR) - 2\log(\gamma).
\end{equation} The lower and upper bounds on $\gamma$ is given by
\begin{eqnarray}
1\leq \gamma \leq 1+\left\|\frac{h_{21}}{h_{22}\SNR^{(1-\rho)/2}-h\SNR^{(\rho-1)/2}}\right\|
\end{eqnarray} where $\gamma$ converges to a  const as $\SNR \rightarrow \infty$. Finally, the achievable GDoF of the primary transmitter is derived as
\begin{equation}\label{eq:GDoF2}
d_{2}(\rho)  \geq \lim_{\SNR,\INR \rightarrow \infty}\frac{R_{2}(\beta^{*})}{\log\SNR}= \rho.
\end{equation}From (\ref{eq:GDoF1}) and (\ref{eq:GDoF2}), the achievable sum GDoF is $1+\rho$.

\subsection{Converse}

For given rates $R_{1}$ and $R_{2}$, we define $R_{sym} = \min\{R_{1},R_{2}\}$ and $\tilde{R} = \max\{R_{1},R_{2}\} - R_{sym}$.  If $R_{1} > R_{2}$ then $W_{1}=(M_{1},\tilde{M})$ and $W_{2}=(M_{2},\zerov)$. In the reverse case, we have that $W_{1}=(M_{1},\zerov)$ and $W_{2}=(M_{2},\tilde{M})$. In both cases, the primary transmitter knows the linear combination, $W_{1}\oplus W_{2} = (M_{1} \oplus M_{2}, \tilde{M})$. From the well-known Crypto Lemma, the $M_{1} \oplus M_{2}$ is mutually statistically independent of $M_{1}$, as well as $M_{1} \oplus M_{2}$ is mutually statistically independent of $M_{2}$.  In this proof, we derive the upper bounds on $R_{sym}$ and $R_{sym}+\tilde{R}$. First, we derive the upper bound on the symmetric rate $R_{sym}$:
\begin{eqnarray*}
n R_{sym} &=& H(M_{1}) = H(M_{1} | M_{1} \oplus M_{2}, \tilde{M})\\
&=& H(M_{1}|M_{1} \oplus M_{2}, \tilde{M})- H(M_{1}|Y_{1}^{n}, M_{1}\oplus M_{2}, \tilde{M})\\
&&+ H(M_{1}| Y_{1}^{n}, M_{1} \oplus M_{2},\tilde{M})\\
&\stackrel{(a)}{\leq}& I(M_{1};Y_{1}^{n}|M_{1} \oplus M_{2},\tilde{M})+n\epsilon_{n}\\
%&=& h(Y_{1}^{n}|M_{1} \oplus M_{2}, \tilde{M}) - h(Y_{1}^{n}|M_{1}, M_{1} \oplus M_{2}, \tilde{M})+n\epsilon_{n}\\
&=& h(Y_{1}^{n}|X_{2}^{n}) - h(Y_{1}^{n}|X_{1}^{n},X_{2}^{n})+n\epsilon_{n}\\
&=& I(X_{1}^{n};Y_{1}^{n}|X_{2}^{n}) + n\epsilon_{n}\\
&\leq& n\log(1+\|h_{11}\|^2\SNR)+n\epsilon_{n}
\end{eqnarray*}where (a) follows from the Fano's inequality and data processing inequality as
\begin{equation*}
H(M_{1}|Y_{1}^{n},M_{1} \oplus M_{2},\tilde{M}) \leq H(M_{1}|Y_{1}^{n})\leq H(M_{1}|\hat{M}_{1})\leq n\epsilon_{n}.
\end{equation*} In the same manner, we get:
\begin{eqnarray*}
n R_{sym} &=& H(M_{2}) = H(M_{2}|M_{1}\oplus M_{2},\tilde{M})\\
%&=& H(M_{2}|M_{1} \oplus M_{2},\tilde{M}) - H(M_{2}|Y_{2}^{n},M_{1} \oplus M_{2},\tilde{M}) + H(M_{2}|Y_{2}^{n}, M_{1}\oplus M_{2},\tilde{M})\\
%&\leq& I(M_{2};Y_{2}^{n}|M_{1} \oplus M_{2},\tilde{M}) + n\epsilon_{n}\\
%&=& h(Y_{2}^{n}|M_{1} \oplus M_{2},\tilde{M}) - h(Y_{2}^{n}|M_{1} \oplus M_{2}, M_{2},\tilde{M}) + n\epsilon_{n}\\
%&=& h(Y_{2}^{n}|X_{2}^{n}) - h(Y_{2}^{n}|X_{1}^{n},X_{2}^{n}) + n \epsilon_{n}\\
%&=& I(X_{1}^{n};Y_{2}^{n}|X_{2}^{n}) + n\epsilon_{n}\\
&\leq& n\log(1+\|h_{21}\|^2\INR) + n\epsilon_{n}.
\end{eqnarray*} From the above, we have the upper bound on $R_{sym}$ as
\begin{equation}\label{eq:upper1}
R_{sym} \leq \min\{\log(1+\|h_{11}\|^2\SNR), \log(1+\|h_{21}\|^2\INR)\}.
\end{equation} The upper bound on $R_{\ell}$ can be computed as
\begin{eqnarray*}
n R_{\ell} &\leq& H(W_{\ell})\\
&=& H(W_{\ell}) - H(W_{\ell}|Y_{\ell}^{n}) + H(W_{\ell}|Y_{\ell}^{n})\\
&\leq& I(W_{\ell} ; Y_{\ell}^{n}) + n\epsilon_{n}\\
%&=& h(Y_{\ell}^{n}) - h(Y_{\ell}^{n}|W_{\ell})+n\epsilon_{n}\\
&\leq& h(Y_{\ell}^{n}) - h(Y_{\ell}^{n}|W_{1},W_{2})+n\epsilon_{n}\\
&=& I(X_{1}^{n},X_{2}^{n};Y_{\ell}^{n}) + \epsilon_{n}\\
&\leq& n\log(1+\|h_{\ell 1}\|^2\SNR + \|h_{\ell 2}\|^2\INR) +n\epsilon_{n}.
\end{eqnarray*}Since $R_{sym} + \tilde{R} = \max\{R_{1},R_{2}\}$, we have:
\begin{eqnarray}\label{eq:upper2}
R_{sym} + \tilde{R} &\leq& \max\{\log(1+\|h_{11}\|^2\SNR+\|h_{12}\|^2\INR),\nonumber\\
&&\log(1+\|h_{21}\|^2\SNR+\|h_{22}\|^2\INR)\}.
\end{eqnarray}
Using (\ref{eq:upper1}), (\ref{eq:upper2}), and $\INR = \SNR^\rho$, we have the upper bounds in the asymptotic case:
\begin{eqnarray*}
\lim_{\SNR \rightarrow \infty}\Big(\frac{R_{sym}}{\log\SNR} + \frac{R_{sym} + \tilde{R}}{\log\SNR}\Big)&\leq& \min\{1,\rho\}+\max\{1,\rho\}.
\end{eqnarray*} Finally we have the upper bound on the sum GDoF as
\begin{equation}
d_{\mbox{\tiny{sum}}}=\lim_{\SNR \rightarrow \infty} \frac{2R_{sym} + \tilde{R}}{\log\SNR} \leq 1 + \rho.
\end{equation} This completes the proof.

%% Or manual references (pay attention to consistency!):


\begin{thebibliography}{1}
%\bibitem{Kramer} G. Kramer, ``Review of rate regions for interference channels," in {\em proceedings of International Zurich Seminar on Communications}, Zurich, Switzerland, Feb. 2006.
%
%
%
%\bibitem{Telatar} E. Telatar and D. Tse, ``Bounds on the capacity region of a class of interference channel," in {\em proceedings of IEEE International Symposium on Information Theory (ISIT)}, Nice, France, Jun 2007.
%

%\bibitem{Flanagan} T. Flanagan, ``Creating cloud base stations with TI's KeyStone multicore architecture," \emph{Texas Instruments White Paper} Oct., 2011.

%\bibitem{Lin} Y. Lin, L. Shao, Z. Zhu, Q. Wang and R. K. Sabhikhi, ``Wireless network cloud: Architecture and system requirements,"
%\emph{IBM Journal of Research and Development}, vol. 54, pp. 4:1 - 4:12, 2010.

%\bibitem{Marict} I. Marict', B. Bostjancic, and A. Goldsmith, ``Resource allocation for constrained backhaul in picocell networks," in \emph{Proceedings of Information Theory and Application Workshop}, pp. 1-6, La Jolla, CA, USA, Feb. 2011.


%\bibitem{Song-ISIT} S.-N. Hong and G. Caire, ''Reverse Compute and Forward: A Low-Complexity Architecture of Downlink Distributed Antenna System," in {\em proceedings of IEEE International Symposium on Information Theory}, Cambridge, MA, Jul. 2012.

\bibitem{Song-IT} S.-N. Hong and G. Caire, ``Lattice Coding Strategies for Cooperative Distributed Antenna Systems," {\em submitted to IEEE Transactions on Information Theory,} Oct. 2012. [Online] Available:http://arxiv.org/abs/1210.0160.


\bibitem{Maric} I. Maric, R. D. Yates, and G. Kramer, ``Capacity of interference channels with partial transmitter cooperation," {\em IEEE Transactions on Information Theory}, vol. 53, pp. 3536-3548, Oct. 2007.

\bibitem{Wu} W. Wu, S. Vishwanath, and A. Arapostathis, ``Capacity of a class of cognitive radio channels: Interference channels with degraded message sets," {\em IEEE Transactions on Information Theory}, vol. 53, pp. 4391-4399, Nov. 2007.

\bibitem{Jovicic} A. Jovicic and P. Viswanath, ``Cognitive radio: An information-theoretic perspective," {\em IEEE Transactions on Information Theory}, vol. 55, pp. 3945-3958, Sept. 2009.

%\bibitem{Rini} S. Rini, D. Tuninetti, and N. Devroye, ``State of the cognitive interference channel: a new unified inner bound, and capacity to within 1.87 bits," in {\em proceedings of International Zurich Seminar on Communications}, 2010.

\bibitem{Rini} S. Rini, D. Tuninetti, and N. Devroye, ``The Capacity Region of Gaussian Cognitive Radio Channels to within 1.87 bits," in {\em proceedings of IEEE Information Theory Workshop (ITW)}, Cairo, Egypt, Jan. 2010.

\bibitem{Ford} L. R. Ford and D. R. Fulkerson, ``Maximal flow through a network," {\em Canadian Journal of  Mathematics}, vol. 8, pp. 399-404, 1956.
\bibitem{Ahlswede} R. Ahlswede, N. Cai, S.-Y. R. Li, and R. W. Yeung, ``Network information flow," {\em IEEE Transactions on Information Theory,} vol. 49, pp. 371-381, Feb. 2003.

\bibitem{Ho} T. Ho, M. Medard, R. Koetter, D. R. Karger, M. Effros, J. Shi, and B. Leong, ``A Random Linear Network Coding Approach to Multicast," {\em IEEE Transactions on Information Theory}, vol. 52, pp. 4413-4430, Oct. 2006.

\bibitem{Etkin} R. Etkin, D. N. C. Tse, and H. Wang, ``Gaussian interference channel capacity to within one bit," {\em IEEE Transactions on information theory}, vol. 54, pp. 5534-5562, Dec. 2008.

\bibitem{Nazer2011} B. Nazer and M. Gastpar, ``Compute-and-Forward: Harnessing Interference through Structured Codes,''\emph{IEEE Transactions on Information Theory}, vol. 57, pp. 6463-6486, Oct. 2011.


\bibitem{Erez2004} U. Erez and R. Zamir, ``Achieving $\frac{1}{2}\log(1+\SNR)$ on the AWGN channel with lattice encoding and decoding," \emph{IEEE Transactions on Information Theory}, vol. 50, pp. 2293-2314, Oct. 2004.

\bibitem{Zamir} R. Zamir, S. Shamai, and U. Erez, ``Nested Linear/Lattice Codes for Structured Multiterminal Binning," \emph{IEEE Transactions on Information Theory}, vol. 48, pp. 1250-1276, June, 2002.


%\bibitem{Boyd} S. Boyd and L. Vandenberghe, \emph{Convex Optimization}, Cambridge University Press, 2004.


%\bibitem{Kim} A. El Gamal and Y.-H. Kim, ``Network Information Theory," Cambridge University Press, 2011.

%\bibitem{Chiachi} C. Huang and S. A. Jafer, ``Degrees of Freedom of the MIMO Interference Channel with Cooperation and Cognition," {\em IEEE Transactions on Information Theory}, vol. 55, pp. 4211-4220, Sep. 2009.




%\bibitem{Dobrushin}R. L. Dobrushin, ``Asymptotic optimality of group and systematic codes for some channels," \emph{Theory of Probability and its Applications}, vol. 8, pp. 47-59, 1963.


%\bibitem{Costa} M. H. M. Costa, ``Writing on dirty paper," {\em IEEE Transactions on Information Theory}, vol. IT-29, pp. 439-441, May 1983.

%\bibitem{Hoon} H. Hoon, H. Papadopoulos, and G. Caire, ``MIMO broadcast channel optimization under general linear constraint," in {\em Proceedings of IEEE International Symposium on Information Theory (ISIT)}, Austin, Tx, Jun 2010.



%\bibitem{Forney} G. D. Forney, Jr, ``On the role of MMSE estimation in approaching the information-theoretic limits of linear Gaussian channels: Shannon meets Wiener," in {\em proceedings of Allerton Conference on Communication, Control, and Computing,} Monticello, Il, 2003.

%\bibitem{Hong-IT} S.-N. Hong and G. Caire, ``Generalized Degrees of Freedom for Network-Coded Cognitive Interference Channel," {\em In preparation}.
\end{thebibliography}
\end{document}